\documentclass[aps,prl,showpacs,twocolumn,superscriptaddress,10pt]{revtex4-1}

\usepackage{amsfonts,amssymb,bm}
\usepackage{amsmath}
\usepackage{graphicx}
\usepackage{epsfig}
\usepackage{natbib}
\usepackage{mathrsfs}
\usepackage{color}
        
\begin{document}

\newcommand{\ket}[1]{\left| #1 \right\rangle}
\newcommand{\bra}[1]{\left\langle #1 \right|}
\newcommand{\braket}[2]{\left\langle #1 | #2 \right\rangle}
\newcommand{\braopket}[3]{\bra{#1}#2\ket{#3}}
\newcommand{\proj}[1]{| #1\rangle\!\langle #1 |}
\newcommand{\expect}[1]{\left\langle#1\right\rangle}
\newcommand{\Tr}{\mathrm{Tr}}
\def\Id{1\!\mathrm{l}}
\newcommand{\cM}{\mathcal{M}}
\newcommand{\cR}{\mathcal{R}}
\newcommand{\cE}{\mathcal{E}}
\newcommand{\cL}{\mathcal{L}}
\newcommand{\cl}{l}
\newcommand{\cH}{\mathcal{H}}
\newcommand{\cU}{\mathcal{U}}
\newcommand{\cP}{\mathcal{P}}
\newcommand{\reals}{\mathbb{R}}
\newcommand{\grad}{\nabla}
\newcommand{\rhohat}{\hat{\rho}}
\newcommand{\rhoMLE}{\rhohat_\mathrm{MLE}}
\newcommand{\rhotomo}{\rhohat_\mathrm{tomo}}
\newcommand{\diff}{\mathrm{d}\!}
\newcommand{\pdiff}[2]{\frac{\partial #1}{\partial #2}}
\newcommand{\todo}[1]{\color{red}#1}
\def\FCW{1.0\columnwidth}
\def\HCW{0.55\columnwidth}
\def\TPW{0.33\textwidth}


\title{Adaptive quantum state tomography improves accuracy quadratically}


\author{D.H. Mahler}
\email[]{dmahler@physics.utoronto.ca}
\affiliation{Centre for Quantum Information \& Quantum Control and Institute for Optical Sciences, Dept. of Physics, 60 St. George St., University of Toronto, Toronto, Ontario, Canada M5S 1A7}
\affiliation{Canadian Institute for Advanced Research, Toronto, Ontario M5G1Z8}
\author{Lee A. Rozema}
\author{Ardavan Darabi}
\affiliation{Centre for Quantum Information \& Quantum Control and Institute for Optical Sciences, Dept. of Physics, 60 St. George St., University of Toronto, Toronto, Ontario, Canada M5S 1A7}
\affiliation{Canadian Institute for Advanced Research, Toronto, Ontario M5G1Z8}
\author{Christopher Ferrie}
\affiliation{Center for Quantum Information and Control, University of New Mexico, Albuquerque, New Mexico 87131-0001, USA}
\author{Robin Blume-Kohout}
\affiliation{Sandia National Laboratories, Advanced Device Technologies (01425), Albuquerque, New Mexico 87185, USA}
\author{A.M. Steinberg}
\affiliation{Centre for Quantum Information \& Quantum Control and Institute for Optical Sciences, Dept. of Physics, 60 St. George St., University of Toronto, Toronto, Ontario, Canada M5S 1A7}
\affiliation{Canadian Institute for Advanced Research, Toronto, Ontario M5G1Z8}



\date{\today}

\begin{abstract}
We introduce a simple protocol for adaptive quantum state tomography, which reduces the worst-case infidelity ($1-F(\rhohat,\rho)$) between the estimate and the true state from $O(1/\sqrt{N})$ to $O(1/N)$.  It uses a single adaptation step and just one extra measurement setting.  In a linear optical qubit experiment, we demonstrate a full order of magnitude reduction in infidelity (from $0.1\%$ to $0.01\%$) for a modest number of samples ($N\approx 3\times10^4$).
\end{abstract}


\maketitle


Quantum information processing requires reliable, repeatable preparation and transformation of quantum states.  \emph{Quantum state tomography} is used to identify the density matrix $\rho$ that was prepared by such a process.  No finite ensemble of $N$ samples is sufficient to uniquely identify $\rho$, so we \emph{estimate} it, reporting either a single state $\rhohat$ that is ``close'' to $\rho$ with high probability \cite{HradilPRA97,ParisBook04,RBKNJP10,RBKPRL10,GrossPRL10}, or a confidence region of nonzero radius that contains $\rho$ with high probability \cite{ChristandlPRL12,RBK12}.  Both approaches must accept some inaccuracy (the discrepancy between $\rhohat$ and $\rho$) or imprecision (the diameter of the confidence region).  The universal goal of state tomography is to minimize this discrepancy, which has been quantified with various metrics (e.g., trace norm, fidelity, relative entropy, etc.).  In this paper, we focus on the particularly well-motivated \emph{quantum infidelity},
\begin{equation}
1-F(\rhohat,\rho)=1-\Tr\left(\sqrt{\sqrt{\rho}\rhohat\sqrt{\rho}}\right)^2,
\end{equation}
and show that as $N\rightarrow\infty$, \emph{adaptive} tomography reduces expected infidelity from $O(1/\sqrt{N})$ to $O(1/N)$.

Unlike alternative metrics, $1-F(\rhohat,\rho)$ quantifies an important operational quantity:  \emph{how many copies are required to reliably distinguish $\rhohat$ from $\rho$?}.  Without doing justice to the rich body of research behind this simple statement (e.g., \cite{WoottersPRD81,HelstromBook,FuchsThesis,FuchsIEEE99,CalsamigliaPRA08,AudenaertPRL07}\ldots), we summarize as follows.  The discrepancy between $\rhohat$ and $\rho$ given a \emph{single sample} is well described by the trace distance, $|\rhohat-\rho|_1$.  But tomography (i) requires $N\gg1$ samples; (ii) is used to predict experiments on $N\gg1$ samples; and (iii) yields errors that cannot be detected without $N\gg1$ samples.  So the operationally relevant quantity is $\left|\rhohat^{\otimes N} - \rho^{\otimes N}\right|_1$, which for $N\gg1$ behaves as $1-e^{-D(\rhohat,\rho)N}$.  The exponent $D$ is the \emph{quantum Chernoff bound} \cite{AudenaertPRL07}, and $N\approx D\log(1/\epsilon)$ samples are necessary and sufficient to distinguish $\rho$ from $\rhohat$ with confidence $1-\epsilon$.  $D$ is tightly bounded by the logarithm of the fidelity (see \cite{CalsamigliaPRA08}, Eq. 28); when $1-F(\rhohat,\rho) \ll 1$ (which should always be true in tomography!), $-\log(F)\approx 1-F$ and
\begin{equation}
\frac{1-F}{2} \leq D \leq 1-F. \label{eq:Chernoffboundbound}
\end{equation}
Thus, $1-F$ really does (almost uniquely) quantify tomographic inaccuracy; $N \approx [1-F(\rhohat,\rho)]^{-1}$ samples are (up to a factor of 2) necessary and sufficient \footnote{Remarkably, for large $N$, local measurements can discriminate almost as well as joint measurements on all $N$ samples.  If $D_Q$ and $D_C$ are the optimal error exponents for joint and local measurements (respectively), then $(1-F)/2 \leq D_C \leq D_Q \leq 1-F$ \cite{CalsamigliaPRA08}.} to falsify $\rhohat$.  In contrast, Hilbert-Schmidt- and trace-distance have no such $N$-sample meaning, and give wildly misleading metrics of tomographic error.

We show that standard tomography with static measurements can't beat $1-F = O(1/\sqrt{N})$ as $N\to\infty$ for a large and important class of states, then introduce and explain a simple adaptive protocol that achieves $1-F = O(1/N)$ for every state.  Finally, we demonstrate this effect in a linear optical experiment, achieving a 10-fold improvement in infidelity (from 0.1\% to 0.01\% with $N=3\times 10^4$ measurements) over standard tomography.  We believe this protocol will have wide application, particularly in situations where the rate of data collection is small, such as post-selected optical systems (e.g. \cite{JWP2011}, where data were collected at approximately $9$ measurements per hour).

Adaptivity has been proposed in various contexts.  Single-step adaptive tomography was first analyzed by \cite{GillPRA00}, then refined in \cite{BaganPRA06,BaganPRL06,HuszarPRA12}.  A scheme similar to ours (and its efficacy for \emph{pure} states) was analyzed in \cite{RehacekPRA04}.  Ref. \cite{OkamotoPRL12} recently treated state estimation as parameter estimation, obtaining results complementary, but largely orthogonal, to those reported here.  Here, we present both an experimental demonstration and simple, self-contained derivation of: (1) why quantum fidelity is significant; (2) why adaptive tomography achieves far better infidelity; and (3) how the adaptation should be done.  We optimize \emph{worst-case} infidelity over all states, not just pure states \cite{RehacekPRA04} or specific ensembles of mixed states (e.g. Ref. \cite{BaganPRL06} achieved high average fidelity, but low fidelity on nearly-pure states).

\section{Adaptive tomography}

Static tomography uses data from a fixed set of measurements.  Different measurements yield subtly different tomographic accuracy \cite{DeBurghPRA08}, but to leading order, ``good'' protocols for single-qubit tomography provide equal information \cite{ScottJPA06} about every component of the unknown density matrix $\rho$,
\begin{equation}
\rho = \frac12\left(\Id + \expect{\sigma_x}\sigma_x + \expect{\sigma_y}\sigma_y + \expect{\sigma_z}\sigma_z\right). \label{eq:xyz}
\end{equation}
The canonical example involves measuring the three Pauli operators ($\sigma_x$, $\sigma_y$, $\sigma_z$).  This minimizes the variance of the estimator $\rhohat$ -- but not the expected infidelity, for two reasons.

\begin{figure}[ht]
\includegraphics[width=\FCW]{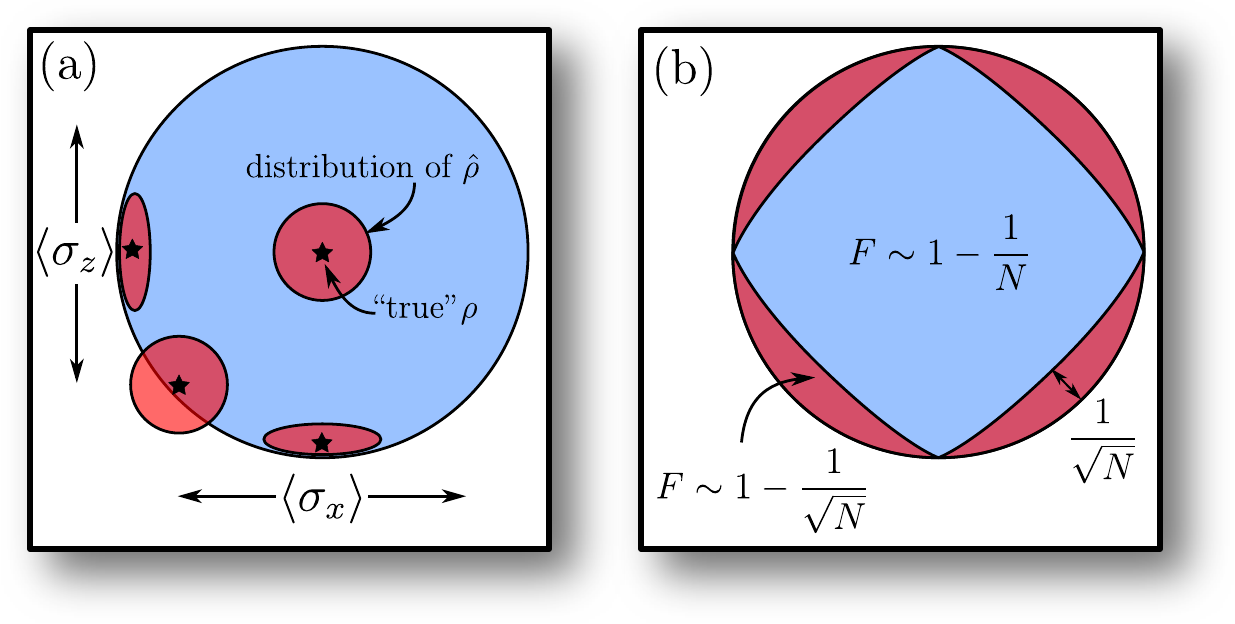}
\caption{Two features of qubit tomography with Pauli measurements (shown for an equatorial cross-section of the Bloch sphere): \textbf{(a)} The distribution or ``scatter'' of any unbiased estimator $\rhohat$ (depicted by dull red ellipses) varies with the true state $\rho$ (black stars at the center of ellipses); \textbf{(b)} The expected infidelity between $\rhohat$ and $\rho$ as a function of $\rho$.   Within the Bloch sphere, the expected infidelity is $O\left(1/N\right)$.  But in a thin shell of nearly-pure states (of thickness $O\left(1/\sqrt{N}\right)$), it scales as $O\left(1/\sqrt{N}\right)$ -- \emph{except} when $\rho$ is aligned with a measurement axis (Pauli $X$, $Y$, or $Z$).\label{fig1} }
\end{figure}

First, \emph{the variance of the estimate $\rhohat$ depends also on $\rho$ itself}.  Consider the linear inversion estimator $\rhohat_{\mathrm{lin}}$, defined by estimating $\expect{\sigma_z} = \frac{n_\uparrow - n_\downarrow}{n_\uparrow + n_\downarrow}$ (and similarly for $\expect{\sigma_x}$ and $\expect{\sigma_y}$), and substituting into Eq. \ref{eq:xyz}.  Each measurement behaves like $N/3$ flips of a coin with bias $p_k = \frac12(1+\expect{\sigma_k})$, and yields
\begin{eqnarray}
\hat{p_k} &=& p_k \pm \sqrt{\frac{3}{N}}\sqrt{p_k(1-p_k)} \label{eq:pvariance}\\
\Rightarrow \expect{\sigma_k}_{\mathrm{estimated}} &=& \expect{\sigma_k}_{\mathrm{true}} \pm \sqrt{\frac{3}{2N}}\sqrt{1-\expect{\sigma_k}^2}. \label{eq:variance}
\end{eqnarray}
When $\expect{\sigma_k}\approx 0$, its estimate has a large variance -- but when $\expect{\sigma_k}\approx\pm1$, the variance is very small.  As a result, the variance of $\rhohat$ around $\rho$ is anisotropic and $\rho$-dependent (see Fig. \ref{fig1}a).

Second, \emph{the dependence of infidelity on the error, $\Delta = \rhohat-\rho$, also varies with $\rho$}.  Infidelity is hypersensitive to misestimation of small eigenvalues.  A Taylor expansion of $1-F(\rhohat,\rho)$ yields (in terms of $\rho$'s eigenbasis $\{\ket{i}\}$),
\begin{equation}
1-F(\rho,\rho+\epsilon\Delta) = \frac14\sum_{i,j}{\frac{\braopket{i}{\Delta}{j}^2}{\braopket{i}{\rho}{i}+\braopket{j}{\rho}{j}}} + O(\Delta^3). \label{eq:Taylor}
\end{equation}
Infidelity is quadratic in $\Delta$ -- except that as an eigenvalue $\braopket{i}{\rho}{i}$ approaches $0$, its sensitivity to $\braopket{i}{\Delta}{i}$ diverges; $1-F$ becomes \emph{linear} \footnote{Because $\rho$ lies on the state-set's boundary, the gradient of $F$ need not vanish in order for $\rhohat=\rho$ to be a local maximum.} in $\Delta$:
\begin{equation}
1-F(\rho,\rho+\epsilon\Delta) = \epsilon\sum_{i:\ \braopket{i}{\rho}{i}=0}{\braopket{i}{\Delta}{i}} + O(\Delta^2).
\end{equation}
To minimize infidelity, we must accurately estimate the small eigenvalues of $\rho$, particularly those that are (or appear to be) zero.  For states deep within the Bloch sphere, static tomography achieves infidelity of $O(1/N)$ \cite{GillPRA00,SugiyamaNJP12}.  Typical errors scale as $|\Delta| = O(1/\sqrt{N})$ (Eq. \ref{eq:variance}), and infidelity scales as $1-F = O(|\Delta|^2)$.  But for states with eigenvalues less than $O(1/\sqrt{N})$, infidelity scales as $O(1/\sqrt{N})$.  Quantum information processing relies on nearly-pure states, so this poor scaling is significant.

To achieve better performance, we observe that if $\rho$ is diagonal in one of the measured bases (e.g., $\sigma_z$), then infidelity \emph{always} scales as $O(1/N)$.  The increased sensitivity of $1-F$ to error in small eigenvalues (Eq. \ref{eq:Taylor}) is precisely canceled by the reduced inaccuracy that accompanies a highly biased measurement-outcome distribution (Eq. \ref{eq:variance}).  This suggests an obvious (if na\"ive) solution:  we should simply ensure that we measure the diagonal basis of $\rho$!

This is unreasonable -- knowing $\rho$ would render tomography pointless.  But we \emph{can} perform standard tomography on $N_0<N$ samples, get a preliminary estimate $\rhohat_0$, and measure the remaining $N-N_0$ samples so that one basis  diagonalizes $\rhohat_0$.  This measurement will not diagonalize $\rho$ exactly, but if $N_0\gg1$ it will be fairly close. The angle $\theta$ between the eigenbases of $\rho$ and $\rhohat_0$ is $O(|\Delta|) = O(1/\sqrt{N_0})$.  This implies that if $\rho$ has an eigenvector $\ket{\psi_k}$ with eigenvalue $\lambda_k=0$, then corresponding measurement outcome $\proj{\phi_k}$ will have probability at most $p_k = \sin^2\theta \approx \theta^2 = O(1/N_0)$.  Since we make this measurement on $O(N-N_0)$ copies \footnote{The ``$O$'' notation is necessary here because some of the remaining $N-N_0$ copies may be measured in other bases that make up a complete measurement frame.}, the final error in the estimated $\hat{p_k}$ (and therefore in the eigenvalue $\lambda_k$) is $O(1/\sqrt{N_0(N-N_0)})$.  So using a constant fraction $N_0 = \alpha N$ of the available samples for the preliminary estimation should yield $O(1/N)$ infidelity for \emph{all} states.

A similar protocol was suggested in Ref. \cite{BaganPRL06}, but that analysis concluded that $N_0\propto N^p$ for $p\geq \frac23$ would be sufficient.  This works for \emph{average} infidelity over a particular ensemble, but yields $1-F = O(N^{-5/6})$ for almost all nearly-pure states.

\section{Simulation results}

We performed numerical simulations of single-qubit tomography using four different protocols: (1) standard fixed-measurement tomography; (2) adaptive tomography with $N_0 = N^{2/3}$, as proposed in \cite{BaganPRL06}; (3) adaptive tomography with $N_0 = \alpha N$ (for a range of $\alpha$); and (4) ``known basis'' tomography, wherein we cheat by aligning our measurement frame with $\rho$'s eigenbasis (for all $N$ samples).  We simulated many true states $\rho$, but present a representative case: a pure state with $(\langle\sigma_x\rangle,\langle\sigma_y\rangle,\langle\sigma_z\rangle)=(0.5,1/\sqrt{2},0.5)$
\begin{equation} \label{eq:diagstate}
\ket{\nearrow} = \frac{1}{2}\left(\begin{array}{c}
\sqrt{3} \\ 
\frac{1}{\sqrt{3}}-\frac{2i}{\sqrt{6}}\end{array}
\right)
\end{equation}
  Our results are not particularly sensitive to the exact estimator used; we used maximum-likelihood estimation (MLE) with a quadratic approximation to the negative loglikelihood function:
\begin{equation}
\cl(\rho)=-\log\cL(\rho) \approx \sum^{3}_{k=1}\frac{N_k(\Tr[\rho E_k]-f_k)^2}{f_k(1-f_k)},
\end{equation} 
where $f_k = n_k/N_k$ are the observed frequencies of the $+1$ eigenvectors of the three Pauli operators $\sigma_k$, $E_k$ is the corresponding projector, and $N_k$ is the number of samples on which $\sigma_k$ was measured.  Convex optimization (in MATLAB \cite{yalmip}) was used to find $\rhoMLE$.  Results were averaged over many (typically 150) randomly generated measurement records.

\begin{figure}[ht]
\includegraphics[width=\FCW]{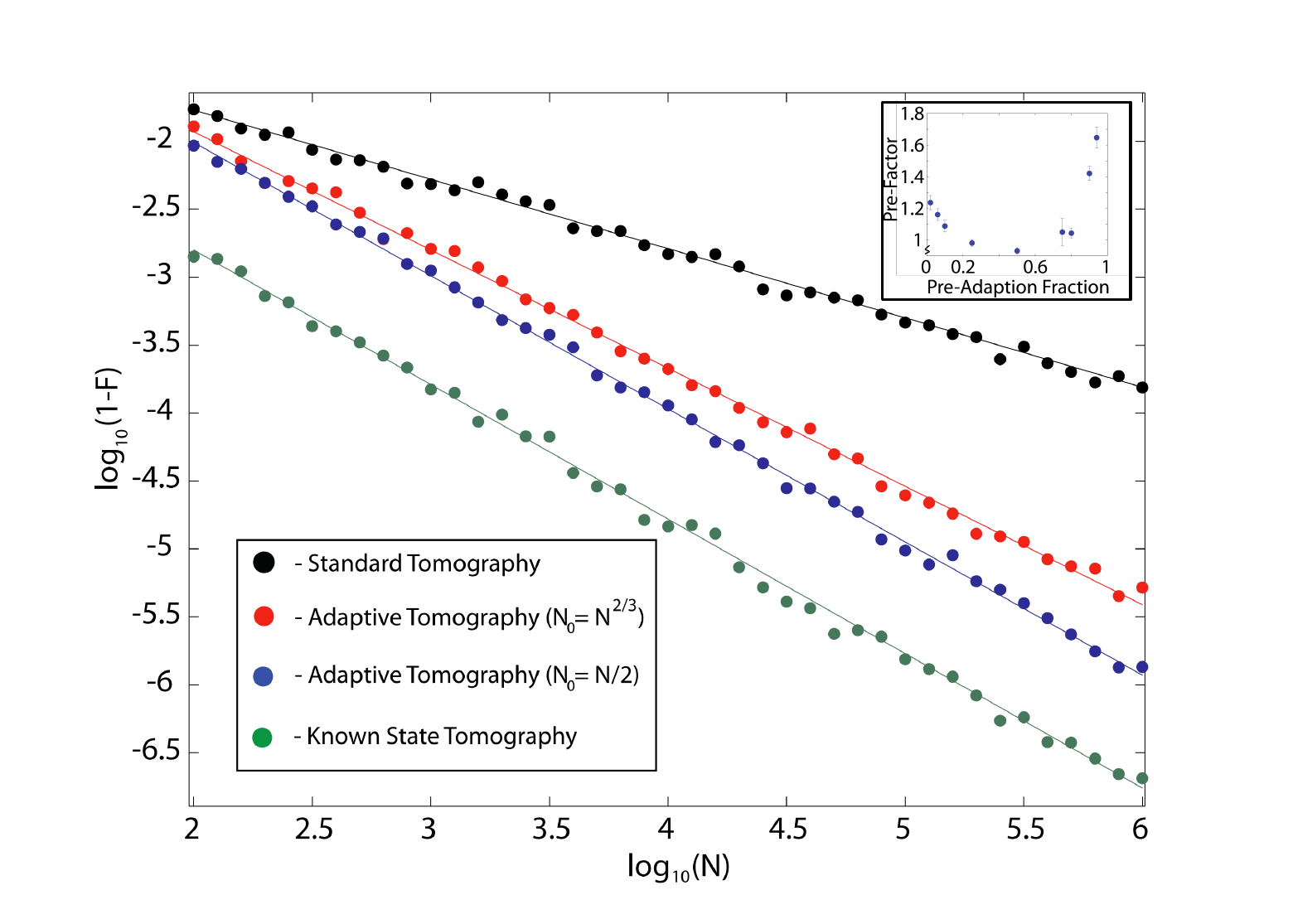}
\caption{Average infidelity $1-F(\rhohat,\rho)$ vs. sample size $N$ for Monte Carlo simulations of four different tomographic procotocols: standard tomography (black), the procedure proposed in \cite{BaganPRL06} using $N_0=N^{2/3}$ (red), our procedure using $N_0 = N/2$ (blue), and ``known basis'' tomography (green).  Both adaptive procedures clearly outperform static tomography, but our procedure clearly outperforms the $N_0=N^{2/3}$ approach, and matches the asymptotic scaling of known-basis tomography.  The inset shows the dependence of the prefactor ($\beta$) on $\alpha=N_0/N$. \label{fig2}}
\end{figure}

Figure \ref{fig2} shows average infidelity versus $N$.  We fit these simulated data to power laws of the form $1-F = \beta N^p$, and found $p=-0.513 \pm 0.006$ (for static tomography), $p=-0.868 \pm 0.008$ (for adaptive tomography with $N_0=N^{2/3}$), $p=-0.980 \pm 0.006$ (for adaptive tomography with $N_0=0.5N$), and $p=-0.993 \pm 0.09$ (for known-basis tomography).  These results are not significantly different \footnote{All quoted uncertainties herein are $1\sigma$, or 68\% confidence intervals.  Therefore, we don't expect the the ``true'' value to lie within the error bars more than 68\% of the time.  Most of the results given here agree with theoretical predictions to within $2\sigma$ (95\% confidence intervals), a common criterion for consistency between data and theory.} from predictions of the simple theory ($p=-\frac12, -\frac56, 1$, and $1$, respectively).   The borderline-significant discrepancy is, we believe, due to boundary effects ($\rhoMLE$ is constrained to be positive).  We also varied $\alpha=N_0/N$ (Fig. \ref{fig2}, inset) and found that $\alpha=\frac12$ optimizes the prefactor ($\beta$).

\section{Experimental results}

\begin{figure}[ht]
\includegraphics[scale=0.3]{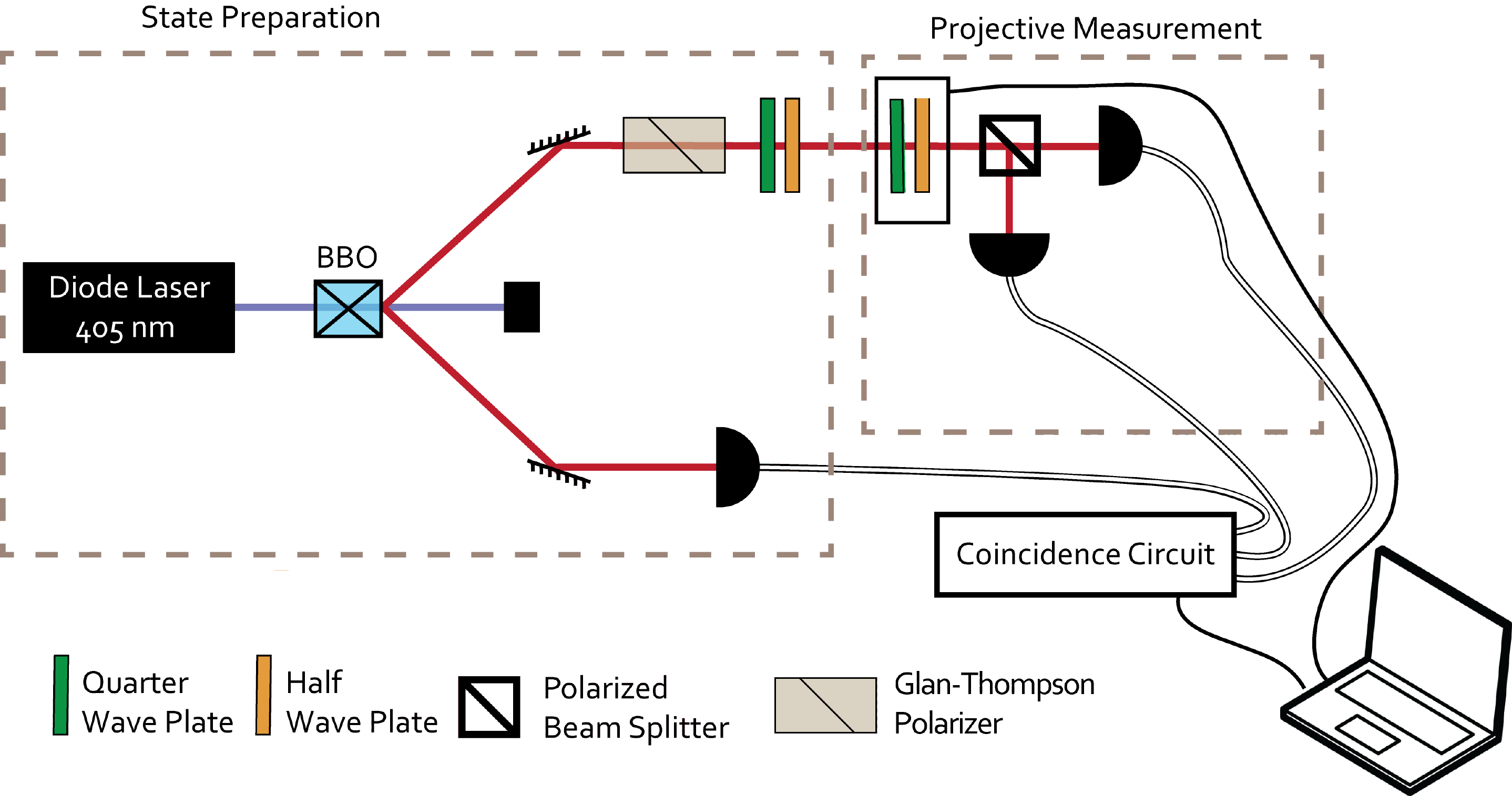}%
\caption{\label{fig3}  Spontaneous parametric downconversion is performed by pumping a nonlinear BBO crystal with linearly polarized light.  One photon is sent directly to a detector as a trigger.  A rotation using a quarter-half waveplate combination prepares the other photon in any desired polarization state. Finally, a projective measurement onto any axis of the Bloch sphere is performed by a quarter-half waveplate combination followed by a polarizing beamsplitter.  The measurement waveplates are connected to a computer to enable adaptation.}
\end{figure}

We implemented our protocol experimentally in linear optics (Fig. \ref{fig3}).  Using type-1 spontaneous parametric down conversion in a nonlinear crystal, photon pairs were created.  One of these photons was sent immediately to a single photon counting module (SPCM) to act as a trigger.  The second photon was sent through a Glan-Thomson polarizer to prepare it in a state of very pure linear polarization.  Computer-controlled waveplates were first used to prepare the polarization state of the photon, and subsequently used in tandem with a polarization beamsplitter to project onto any state on the Bloch sphere.

We compared static and adaptive tomography protocols on a measured state given (in the H/V basis) by
\begin{equation}
\rho = \left(\begin{array}{cc} 0.7711 & 0.2010 + 0.3624i \\ 0.2010 - 0.3624i & 0.2289 \end{array}\right),
\end{equation}
which has purity $\Tr(\rho^2) = 0.991$ and fidelity $F=0.992$ with $\ket{\nearrow}$ (see Eq. \ref{eq:diagstate}).  We identified $\rho$ to within an uncertainty which is at most $O(1/\sqrt{\tilde{N}})$ using one very long ($\tilde{N}=10^7$) static tomography experiment, whose overwhelming size ensures accuracy sufficient to calibrate the other experiments, all of which involve $N\leq3\times 10^4$ photons.

\begin{figure}[ht]
\includegraphics[width=\FCW]{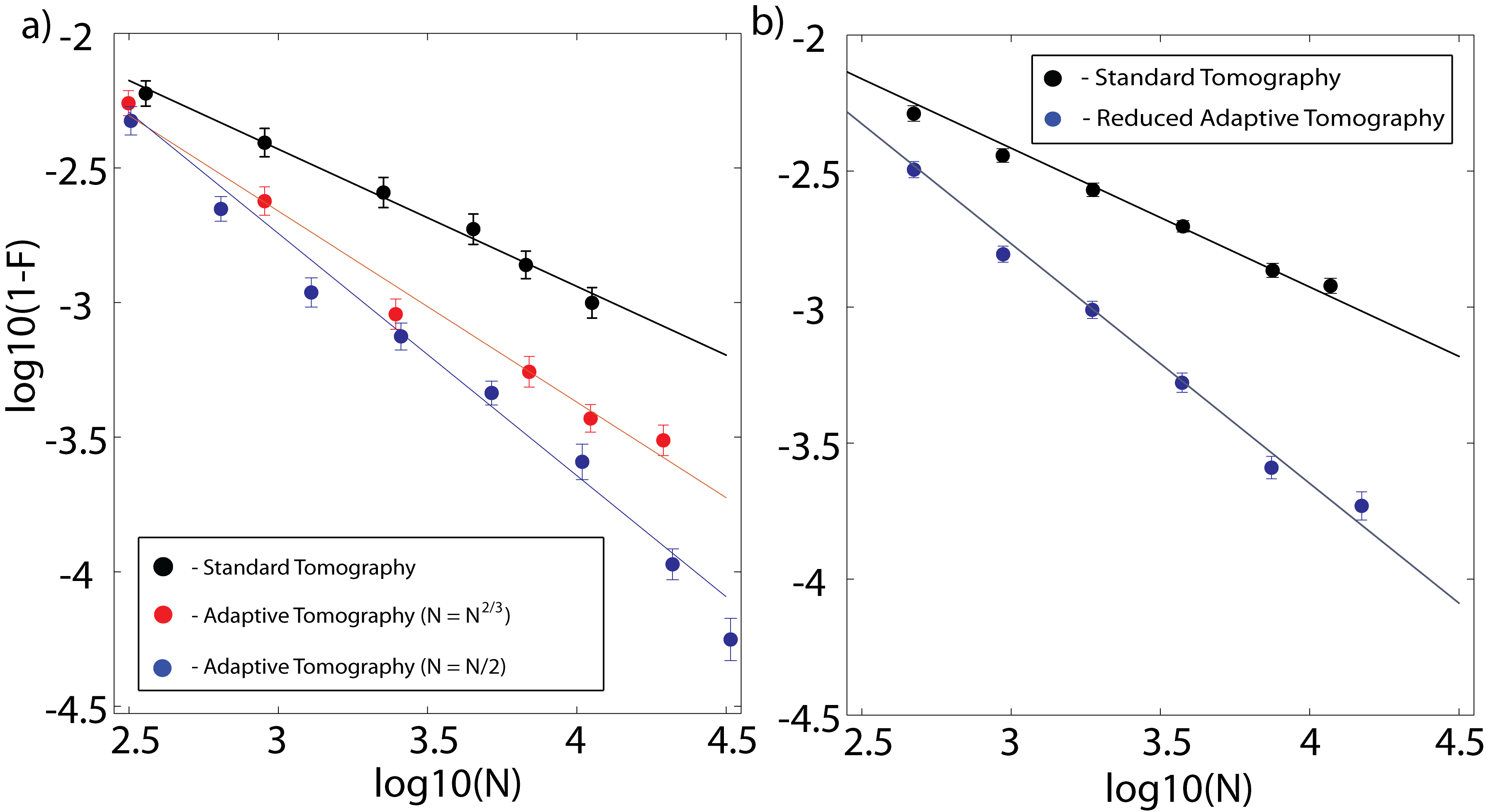}%
\caption{Experimental data: a) The average infidelity $1-F(\rhohat,\rho)$ for the three tomographic protocols shown in Fig. \ref{fig2} vs. the number of samples $N$.  Each average is over 150 different realizations of the experiment. b) Average infidelity $1-F(\rhohat,\rho)$ for standard tomography (black) and \emph{reduced} adaptive tomography (blue) is plotted versus $N$.  Each average is over 200 different realizations of the experiment; error bars are standard deviation of the mean of these samples.  Error bars are standard deviation of the mean of these samples.\label{fig4}  }
\end{figure}

Our ``standard'' (static) protocol involved repeatedly preparing our target state, collecting $N/3$ photons at each of the three measurement settings corresponding to $\sigma_x$, $\sigma_y$, and $\sigma_z$, and computing $\rhoMLE$ as outlined in \cite{JamesPRA01}.  Each data point in figure \ref{fig4}a represents an average over many ($\sim\!150$) repetitions.  

To do adaptive tomography, we measured $N_0=N/2$ photons, used the data to generate an ML estimate $\rhohat_0$, then rotated the measurement bases so that one diagonalized $\rhohat_0$.  So, if the preliminary estimate is
\begin{equation*}
\rhohat_0 = \lambda_1\proj{\psi_1} + \lambda_2\proj{\psi_2},
\end{equation*}
we define $\vert \psi_{3/4} \rangle = (1/2)(\vert\psi_1\rangle \pm \vert \psi_2\rangle)$ and $\vert \psi_{5/6}\rangle = (1/2)(\vert\psi_1\rangle \pm i\vert \psi_2\rangle)$, and then measure the bases $\{\{\vert\psi_1\rangle,\vert \psi_2\rangle\},\{\vert\psi_3\rangle,\vert \psi_4\rangle\},\{\vert\psi_5\rangle,\vert \psi_6\rangle\}\}$.  We measured the remaining $N-N_0$ photons in these new bases and constructed a final ML estimate using the data from both phases.

We fit a power law ($1-F = \beta N^p$) to the average infidelity of each protocol (Fig. \ref{fig4}a), and found $p=-0.51 \pm 0.02$ for standard tomography, $p=-0.71 \pm 0.04$ for the procedure of Ref. \cite{BaganPRL06}, and $p=-0.90 \pm 0.04$ for our adaptive procedure.

Our data generally match the theory; adaptive tomography outperforms standard tomography by an order of magnitude even for modest ($\sim10^4$) $N$.  Experiments that achieve very low infidelities ($\sim10^{-4}$) show small but statistically significant deviations from theory, which we believe can be explained by waveplate misalignment -- fluctuations on the order of $10^{-3}$ radians are sufficient to reproduce the observed deviations in simulations.  For a detailed discussion of systematic error and how it affects our results please see the supplementary material.  

There is an even simpler adaptive procedure.  After obtaining a preliminary estimate $\rhohat_0$, we measured \emph{all} of the remaining $N/2$ samples in the diagonal basis of $\rhohat_0$, neglecting the second and third bases presented in the previous section's protocol.  This \emph{reduced adaptive tomography} procedure requires just one extra measurement setting (full adaptive tomography requires three), but achieves the same $O(\frac{1}{N})$ infidelity (Fig. \ref{fig4}b).  The best fits to the exponent $p$ in $1-F =\beta N^p$ are $p=-0.51 \pm 0.02$ for standard tomography and $p=-0.88 \pm 0.05$ for reduced adaptive tomography (not significantly different from the results shown in Fig. \ref{fig4}a).  In higher dimensional systems, reduced adaptive tomography should provide even greater efficiency advantages.

\section{Discussion}

We demonstrated two easily implemented adaptive tomography procedures that achieve $1-F(\rhohat,\rho) = O(1/N)$ for \emph{every} qubit state.  In contrast, any static tomography protocol will yield infidelity $O(1/\sqrt{N})$ for most nearly-pure states.  Our simplest procedure requires only one additional measurement setting than standard tomography.  We see almost no reason not to use reduced adaptive tomography in future experiments.

Previous work \cite{BaganPRL06} optimized average fidelity over Bures measure, a very respectable choice \cite{HubnerPLA92,PetzJMP96,ZyczkowskiPRA05}.  Unfortunately, the ``hard-to-estimate'' states lie in a thin shell at the surface of the Bloch sphere, whose Bures measure vanishes as $N\to\infty$.  So although the scheme with $N_0\propto N^{2/3}$ proposed in \cite{BaganPRL06} achieves Bures-average infidelity $O(1/N)$, it achieves only $O(1/N^{5/6})$ infidelity for nearly all of the (important) nearly-pure states \footnote{Ironically, restricting the problem to pure states falsely trivializes it -- the average \emph{and} worst-case infidelity is $O(1/N)$ even for static tomography!  The difficulty is not in estimating which pure state we have, but in distinguishing between small eigenvalues ($\lambda=0$ vs $\lambda=1/\sqrt{N}$).}

The $O(1/N)$ infidelity scaling achieved by our scheme is optimal, but the constant can surely be improved -- i.e., if our scheme has asymptotic error $\alpha/N$, a more sophisticated scheme can achieve $\alpha'/N$ with $\alpha'<\alpha$.  The absolutely optimal protocol requires joint measurements on all $N$ samples \cite{MassarPRL95}, and will outperform any local measurement.  There is undoubtedly some marginal benefit to adapting more than once, but we have shown that a single adaptation is sufficient to achieve $O(1/N)$ scaling.


\begin{acknowledgments}
DHM, LAR, AD, and AMS thank NSERC and CIFAR for support, and Alan Stummer for designing the coincidence circuit.  CF was supported in part by NSF Grant Nos.~PHY-1212445 and PHY-1005540 and an NSERC PDF.  RBK was supported by the LDRD program at Sandia National Laboratories, a multi-program laboratory operated by Sandia Corporation, a wholly owned subsidiary of Lockheed Martin Corporation, for the U.S. Department of Energy's National Nuclear Security Administration under contract DE-AC04-94AL85000. 
\end{acknowledgments}

\bibliography{../Abib}{}

\begin{thebibliography}{34}%
\makeatletter
\providecommand \@ifxundefined [1]{%
 \@ifx{#1\undefined}
}%
\providecommand \@ifnum [1]{%
 \ifnum #1\expandafter \@firstoftwo
 \else \expandafter \@secondoftwo
 \fi
}%
\providecommand \@ifx [1]{%
 \ifx #1\expandafter \@firstoftwo
 \else \expandafter \@secondoftwo
 \fi
}%
\providecommand \natexlab [1]{#1}%
\providecommand \enquote  [1]{``#1''}%
\providecommand \bibnamefont  [1]{#1}%
\providecommand \bibfnamefont [1]{#1}%
\providecommand \citenamefont [1]{#1}%
\providecommand \href@noop [0]{\@secondoftwo}%
\providecommand \href [0]{\begingroup \@sanitize@url \@href}%
\providecommand \@href[1]{\@@startlink{#1}\@@href}%
\providecommand \@@href[1]{\endgroup#1\@@endlink}%
\providecommand \@sanitize@url [0]{\catcode `\\12\catcode `\$12\catcode
  `\&12\catcode `\#12\catcode `\^12\catcode `\_12\catcode `\%12\relax}%
\providecommand \@@startlink[1]{}%
\providecommand \@@endlink[0]{}%
\providecommand \url  [0]{\begingroup\@sanitize@url \@url }%
\providecommand \@url [1]{\endgroup\@href {#1}{\urlprefix }}%
\providecommand \urlprefix  [0]{URL }%
\providecommand \Eprint [0]{\href }%
\providecommand \doibase [0]{http://dx.doi.org/}%
\providecommand \selectlanguage [0]{\@gobble}%
\providecommand \bibinfo  [0]{\@secondoftwo}%
\providecommand \bibfield  [0]{\@secondoftwo}%
\providecommand \translation [1]{[#1]}%
\providecommand \BibitemOpen [0]{}%
\providecommand \bibitemStop [0]{}%
\providecommand \bibitemNoStop [0]{.\EOS\space}%
\providecommand \EOS [0]{\spacefactor3000\relax}%
\providecommand \BibitemShut  [1]{\csname bibitem#1\endcsname}%
\let\auto@bib@innerbib\@empty
\bibitem [{\citenamefont {Hradil}(1997)}]{HradilPRA97}%
  \BibitemOpen
  \bibfield  {author} {\bibinfo {author} {\bibfnamefont {Z.}~\bibnamefont
  {Hradil}},\ }\href@noop {} {\bibfield  {journal} {\bibinfo  {journal} {Phys.
  Rev. A}\ }\textbf {\bibinfo {volume} {55}},\ \bibinfo {pages} {1561}
  (\bibinfo {year} {1997})}\BibitemShut {NoStop}%
\bibitem [{\citenamefont {Paris}\ and\ \citenamefont
  {Rehacek}(2004)}]{ParisBook04}%
  \BibitemOpen
  \bibfield  {author} {\bibinfo {author} {\bibfnamefont {M.}~\bibnamefont
  {Paris}}\ and\ \bibinfo {author} {\bibfnamefont {J.}~\bibnamefont
  {Rehacek}},\ }\href@noop {} {\emph {\bibinfo {title} {Quantum state
  estimation}}},\ Vol.\ \bibinfo {volume} {649}\ (\bibinfo  {publisher}
  {Springer},\ \bibinfo {year} {2004})\BibitemShut {NoStop}%
\bibitem [{\citenamefont {Blume-Kohout}(2010{\natexlab{a}})}]{RBKNJP10}%
  \BibitemOpen
  \bibfield  {author} {\bibinfo {author} {\bibfnamefont {R.}~\bibnamefont
  {Blume-Kohout}},\ }\href@noop {} {\bibfield  {journal} {\bibinfo  {journal}
  {N. J. Phys.}\ }\textbf {\bibinfo {volume} {12}},\ \bibinfo {pages} {043034}
  (\bibinfo {year} {2010}{\natexlab{a}})}\BibitemShut {NoStop}%
\bibitem [{\citenamefont {Blume-Kohout}(2010{\natexlab{b}})}]{RBKPRL10}%
  \BibitemOpen
  \bibfield  {author} {\bibinfo {author} {\bibfnamefont {R.}~\bibnamefont
  {Blume-Kohout}},\ }\href@noop {} {\bibfield  {journal} {\bibinfo  {journal}
  {Phys. Rev. Lett.}\ }\textbf {\bibinfo {volume} {105}},\ \bibinfo {pages}
  {200504} (\bibinfo {year} {2010}{\natexlab{b}})}\BibitemShut {NoStop}%
\bibitem [{\citenamefont {Gross}\ \emph {et~al.}(2010)\citenamefont {Gross},
  \citenamefont {Liu}, \citenamefont {Flammia}, \citenamefont {Becker},\ and\
  \citenamefont {Eisert}}]{GrossPRL10}%
  \BibitemOpen
  \bibfield  {author} {\bibinfo {author} {\bibfnamefont {D.}~\bibnamefont
  {Gross}}, \bibinfo {author} {\bibfnamefont {Y.-K.}\ \bibnamefont {Liu}},
  \bibinfo {author} {\bibfnamefont {S.~T.}\ \bibnamefont {Flammia}}, \bibinfo
  {author} {\bibfnamefont {S.}~\bibnamefont {Becker}}, \ and\ \bibinfo {author}
  {\bibfnamefont {J.}~\bibnamefont {Eisert}},\ }\href@noop {} {\bibfield
  {journal} {\bibinfo  {journal} {Phys. Rev. Lett.}\ }\textbf {\bibinfo
  {volume} {105}},\ \bibinfo {pages} {150401} (\bibinfo {year}
  {2010})}\BibitemShut {NoStop}%
\bibitem [{\citenamefont {Christandl}\ and\ \citenamefont
  {Renner}(2012)}]{ChristandlPRL12}%
  \BibitemOpen
  \bibfield  {author} {\bibinfo {author} {\bibfnamefont {M.}~\bibnamefont
  {Christandl}}\ and\ \bibinfo {author} {\bibfnamefont {R.}~\bibnamefont
  {Renner}},\ }\href@noop {} {\bibfield  {journal} {\bibinfo  {journal} {Phys.
  Rev. Lett.}\ }\textbf {\bibinfo {volume} {109}},\ \bibinfo {pages} {120403}
  (\bibinfo {year} {2012})}\BibitemShut {NoStop}%
\bibitem [{\citenamefont {Blume-Kohout}(2012)}]{RBK12}%
  \BibitemOpen
  \bibfield  {author} {\bibinfo {author} {\bibfnamefont {R.}~\bibnamefont
  {Blume-Kohout}},\ }\href@noop {} {\bibfield  {journal} {\bibinfo  {journal}
  {arXiv:1202.5270}\ } (\bibinfo {year} {2012})}\BibitemShut {NoStop}%
\bibitem [{\citenamefont {Wootters}(1981)}]{WoottersPRD81}%
  \BibitemOpen
  \bibfield  {author} {\bibinfo {author} {\bibfnamefont {W.~K.}\ \bibnamefont
  {Wootters}},\ }\href {\doibase 10.1103/PhysRevD.23.357} {\bibfield  {journal}
  {\bibinfo  {journal} {Phys. Rev. D}\ }\textbf {\bibinfo {volume} {23}},\
  \bibinfo {pages} {357} (\bibinfo {year} {1981})}\BibitemShut {NoStop}%
\bibitem [{\citenamefont {Helstrom}(1976)}]{HelstromBook}%
  \BibitemOpen
  \bibfield  {author} {\bibinfo {author} {\bibfnamefont {C.~W.}\ \bibnamefont
  {Helstrom}},\ }\href@noop {} {\emph {\bibinfo {title} {Quantum detection and
  estimation theory}}},\ Vol.~\bibinfo {volume} {84}\ (\bibinfo  {publisher}
  {Academic press New York},\ \bibinfo {year} {1976})\BibitemShut {NoStop}%
\bibitem [{\citenamefont {Fuchs}(1996)}]{FuchsThesis}%
  \BibitemOpen
  \bibfield  {author} {\bibinfo {author} {\bibfnamefont {C.~A.}\ \bibnamefont
  {Fuchs}},\ }\href@noop {} {\bibfield  {journal} {\bibinfo  {journal} {arXiv
  preprint quant-ph/9601020}\ } (\bibinfo {year} {1996})}\BibitemShut {NoStop}%
\bibitem [{\citenamefont {Fuchs}\ and\ \citenamefont {van~de
  Graaf}(1999)}]{FuchsIEEE99}%
  \BibitemOpen
  \bibfield  {author} {\bibinfo {author} {\bibfnamefont {C.}~\bibnamefont
  {Fuchs}}\ and\ \bibinfo {author} {\bibfnamefont {J.}~\bibnamefont {van~de
  Graaf}},\ }\href@noop {} {\bibfield  {journal} {\bibinfo  {journal}
  {Information Theory, IEEE Transactions on}\ }\textbf {\bibinfo {volume}
  {45}},\ \bibinfo {pages} {1216} (\bibinfo {year} {1999})}\BibitemShut
  {NoStop}%
\bibitem [{\citenamefont {Calsamiglia}\ \emph {et~al.}(2008)\citenamefont
  {Calsamiglia}, \citenamefont {Mu\~noz Tapia}, \citenamefont {Masanes},
  \citenamefont {Acin},\ and\ \citenamefont {Bagan}}]{CalsamigliaPRA08}%
  \BibitemOpen
  \bibfield  {author} {\bibinfo {author} {\bibfnamefont {J.}~\bibnamefont
  {Calsamiglia}}, \bibinfo {author} {\bibfnamefont {R.}~\bibnamefont {Mu\~noz
  Tapia}}, \bibinfo {author} {\bibfnamefont {L.}~\bibnamefont {Masanes}},
  \bibinfo {author} {\bibfnamefont {A.}~\bibnamefont {Acin}}, \ and\ \bibinfo
  {author} {\bibfnamefont {E.}~\bibnamefont {Bagan}},\ }\href {\doibase
  10.1103/PhysRevA.77.032311} {\bibfield  {journal} {\bibinfo  {journal} {Phys.
  Rev. A}\ }\textbf {\bibinfo {volume} {77}},\ \bibinfo {pages} {032311}
  (\bibinfo {year} {2008})}\BibitemShut {NoStop}%
\bibitem [{\citenamefont {Audenaert}\ \emph {et~al.}(2007)\citenamefont
  {Audenaert}, \citenamefont {Calsamiglia}, \citenamefont {Mu\~noz Tapia},
  \citenamefont {Bagan}, \citenamefont {Masanes}, \citenamefont {Acin},\ and\
  \citenamefont {Verstraete}}]{AudenaertPRL07}%
  \BibitemOpen
  \bibfield  {author} {\bibinfo {author} {\bibfnamefont {K.~M.~R.}\
  \bibnamefont {Audenaert}}, \bibinfo {author} {\bibfnamefont {J.}~\bibnamefont
  {Calsamiglia}}, \bibinfo {author} {\bibfnamefont {R.}~\bibnamefont {Mu\~noz
  Tapia}}, \bibinfo {author} {\bibfnamefont {E.}~\bibnamefont {Bagan}},
  \bibinfo {author} {\bibfnamefont {L.}~\bibnamefont {Masanes}}, \bibinfo
  {author} {\bibfnamefont {A.}~\bibnamefont {Acin}}, \ and\ \bibinfo {author}
  {\bibfnamefont {F.}~\bibnamefont {Verstraete}},\ }\href {\doibase
  10.1103/PhysRevLett.98.160501} {\bibfield  {journal} {\bibinfo  {journal}
  {Phys. Rev. Lett.}\ }\textbf {\bibinfo {volume} {98}},\ \bibinfo {pages}
  {160501} (\bibinfo {year} {2007})}\BibitemShut {NoStop}%
\bibitem [{Note1()}]{Note1}%
  \BibitemOpen
  \bibinfo {note} {Remarkably, for large $N$, local measurements can
  discriminate almost as well as joint measurements on all $N$ samples. If
  $D_Q$ and $D_C$ are the optimal error exponents for joint and local
  measurements (respectively), then $(1-F)/2 \leq D_C \leq D_Q \leq 1-F$ \cite
  {CalsamigliaPRA08}.}\BibitemShut {Stop}%
\bibitem [{\citenamefont {Yao}\ \emph {et~al.}(2011)\citenamefont {Yao},
  \citenamefont {Wang}, \citenamefont {Xu}, \citenamefont {Lu}, \citenamefont
  {Pan}, \citenamefont {Bao}, \citenamefont {Peng}, \citenamefont {Lu},
  \citenamefont {Chen},\ and\ \citenamefont {Pan}}]{JWP2011}%
  \BibitemOpen
  \bibfield  {author} {\bibinfo {author} {\bibfnamefont {X.-C.}\ \bibnamefont
  {Yao}}, \bibinfo {author} {\bibfnamefont {T.-X.}\ \bibnamefont {Wang}},
  \bibinfo {author} {\bibfnamefont {P.}~\bibnamefont {Xu}}, \bibinfo {author}
  {\bibfnamefont {H.}~\bibnamefont {Lu}}, \bibinfo {author} {\bibfnamefont
  {G.-S.}\ \bibnamefont {Pan}}, \bibinfo {author} {\bibfnamefont {X.-H.}\
  \bibnamefont {Bao}}, \bibinfo {author} {\bibfnamefont {C.-Z.}\ \bibnamefont
  {Peng}}, \bibinfo {author} {\bibfnamefont {C.-Y.}\ \bibnamefont {Lu}},
  \bibinfo {author} {\bibfnamefont {Y.-A.}\ \bibnamefont {Chen}}, \ and\
  \bibinfo {author} {\bibfnamefont {J.-W.}\ \bibnamefont {Pan}},\ }\href
  {http://dx.doi.org/10.1038/nphoton.2011.354} {\bibfield  {journal} {\bibinfo
  {journal} {Nature Photonics}\ }\textbf {\bibinfo {volume} {6}},\ \bibinfo
  {pages} {225} (\bibinfo {year} {2011})}\BibitemShut {NoStop}%
\bibitem [{\citenamefont {Gill}\ and\ \citenamefont
  {Massar}(2000)}]{GillPRA00}%
  \BibitemOpen
  \bibfield  {author} {\bibinfo {author} {\bibfnamefont {R.~D.}\ \bibnamefont
  {Gill}}\ and\ \bibinfo {author} {\bibfnamefont {S.}~\bibnamefont {Massar}},\
  }\href {\doibase 10.1103/PhysRevA.61.042312} {\bibfield  {journal} {\bibinfo
  {journal} {Phys. Rev. A}\ }\textbf {\bibinfo {volume} {61}},\ \bibinfo
  {pages} {042312} (\bibinfo {year} {2000})}\BibitemShut {NoStop}%
\bibitem [{\citenamefont {Bagan}\ \emph
  {et~al.}(2006{\natexlab{a}})\citenamefont {Bagan}, \citenamefont {Ballester},
  \citenamefont {Gill}, \citenamefont {Monras},\ and\ \citenamefont
  {Munoz-Tapia}}]{BaganPRA06}%
  \BibitemOpen
  \bibfield  {author} {\bibinfo {author} {\bibfnamefont {E.}~\bibnamefont
  {Bagan}}, \bibinfo {author} {\bibfnamefont {M.}~\bibnamefont {Ballester}},
  \bibinfo {author} {\bibfnamefont {R.~D.}\ \bibnamefont {Gill}}, \bibinfo
  {author} {\bibfnamefont {A.}~\bibnamefont {Monras}}, \ and\ \bibinfo {author}
  {\bibfnamefont {R.}~\bibnamefont {Munoz-Tapia}},\ }\href@noop {} {\bibfield
  {journal} {\bibinfo  {journal} {Phys. Rev. A}\ }\textbf {\bibinfo {volume}
  {73}},\ \bibinfo {pages} {032301} (\bibinfo {year}
  {2006}{\natexlab{a}})}\BibitemShut {NoStop}%
\bibitem [{\citenamefont {Bagan}\ \emph
  {et~al.}(2006{\natexlab{b}})\citenamefont {Bagan}, \citenamefont {Ballester},
  \citenamefont {Gill}, \citenamefont {Munoz-Tapia},\ and\ \citenamefont
  {Romero-Isart}}]{BaganPRL06}%
  \BibitemOpen
  \bibfield  {author} {\bibinfo {author} {\bibfnamefont {E.}~\bibnamefont
  {Bagan}}, \bibinfo {author} {\bibfnamefont {M.}~\bibnamefont {Ballester}},
  \bibinfo {author} {\bibfnamefont {R.}~\bibnamefont {Gill}}, \bibinfo {author}
  {\bibfnamefont {R.}~\bibnamefont {Munoz-Tapia}}, \ and\ \bibinfo {author}
  {\bibfnamefont {O.}~\bibnamefont {Romero-Isart}},\ }\href@noop {} {\bibfield
  {journal} {\bibinfo  {journal} {Phys. Rev. Lett.}\ }\textbf {\bibinfo
  {volume} {97}},\ \bibinfo {pages} {130501} (\bibinfo {year}
  {2006}{\natexlab{b}})}\BibitemShut {NoStop}%
\bibitem [{\citenamefont {Husz\'ar}\ and\ \citenamefont
  {Houlsby}(2012)}]{HuszarPRA12}%
  \BibitemOpen
  \bibfield  {author} {\bibinfo {author} {\bibfnamefont {F.}~\bibnamefont
  {Husz\'ar}}\ and\ \bibinfo {author} {\bibfnamefont {N.~M.~T.}\ \bibnamefont
  {Houlsby}},\ }\href {\doibase 10.1103/PhysRevA.85.052120} {\bibfield
  {journal} {\bibinfo  {journal} {Phys. Rev. A}\ }\textbf {\bibinfo {volume}
  {85}},\ \bibinfo {pages} {052120} (\bibinfo {year} {2012})}\BibitemShut
  {NoStop}%
\bibitem [{\citenamefont {\ifmmode \check{R}\else
  \v{R}\fi{}eh\'a\ifmmode~\check{c}\else \v{c}\fi{}ek}\ \emph
  {et~al.}(2004)\citenamefont {\ifmmode \check{R}\else
  \v{R}\fi{}eh\'a\ifmmode~\check{c}\else \v{c}\fi{}ek}, \citenamefont
  {Englert},\ and\ \citenamefont {Kaszlikowski}}]{RehacekPRA04}%
  \BibitemOpen
  \bibfield  {author} {\bibinfo {author} {\bibfnamefont {J.}~\bibnamefont
  {\ifmmode \check{R}\else \v{R}\fi{}eh\'a\ifmmode~\check{c}\else
  \v{c}\fi{}ek}}, \bibinfo {author} {\bibfnamefont {B.-G.}\ \bibnamefont
  {Englert}}, \ and\ \bibinfo {author} {\bibfnamefont {D.}~\bibnamefont
  {Kaszlikowski}},\ }\href {\doibase 10.1103/PhysRevA.70.052321} {\bibfield
  {journal} {\bibinfo  {journal} {Phys. Rev. A}\ }\textbf {\bibinfo {volume}
  {70}},\ \bibinfo {pages} {052321} (\bibinfo {year} {2004})}\BibitemShut
  {NoStop}%
\bibitem [{\citenamefont {Okamoto}\ \emph {et~al.}(2012)\citenamefont
  {Okamoto}, \citenamefont {Iefuji}, \citenamefont {Oyama}, \citenamefont
  {Yamagata}, \citenamefont {Imai}, \citenamefont {Fujiwara},\ and\
  \citenamefont {Takeuchi}}]{OkamotoPRL12}%
  \BibitemOpen
  \bibfield  {author} {\bibinfo {author} {\bibfnamefont {R.}~\bibnamefont
  {Okamoto}}, \bibinfo {author} {\bibfnamefont {M.}~\bibnamefont {Iefuji}},
  \bibinfo {author} {\bibfnamefont {S.}~\bibnamefont {Oyama}}, \bibinfo
  {author} {\bibfnamefont {K.}~\bibnamefont {Yamagata}}, \bibinfo {author}
  {\bibfnamefont {H.}~\bibnamefont {Imai}}, \bibinfo {author} {\bibfnamefont
  {A.}~\bibnamefont {Fujiwara}}, \ and\ \bibinfo {author} {\bibfnamefont
  {S.}~\bibnamefont {Takeuchi}},\ }\href@noop {} {\bibfield  {journal}
  {\bibinfo  {journal} {Phys. Rev. Lett.}\ }\textbf {\bibinfo {volume} {109}},\
  \bibinfo {pages} {130404} (\bibinfo {year} {2012})}\BibitemShut {NoStop}%
\bibitem [{\citenamefont {de~Burgh}\ \emph {et~al.}(2008)\citenamefont
  {de~Burgh}, \citenamefont {Langford}, \citenamefont {Doherty},\ and\
  \citenamefont {Gilchrist}}]{DeBurghPRA08}%
  \BibitemOpen
  \bibfield  {author} {\bibinfo {author} {\bibfnamefont {M.~D.}\ \bibnamefont
  {de~Burgh}}, \bibinfo {author} {\bibfnamefont {N.~K.}\ \bibnamefont
  {Langford}}, \bibinfo {author} {\bibfnamefont {A.~C.}\ \bibnamefont
  {Doherty}}, \ and\ \bibinfo {author} {\bibfnamefont {A.}~\bibnamefont
  {Gilchrist}},\ }\href@noop {} {\bibfield  {journal} {\bibinfo  {journal}
  {Phys. Rev. A}\ }\textbf {\bibinfo {volume} {78}},\ \bibinfo {pages} {052122}
  (\bibinfo {year} {2008})}\BibitemShut {NoStop}%
\bibitem [{\citenamefont {Scott}(2006)}]{ScottJPA06}%
  \BibitemOpen
  \bibfield  {author} {\bibinfo {author} {\bibfnamefont {A.~J.}\ \bibnamefont
  {Scott}},\ }\href@noop {} {\bibfield  {journal} {\bibinfo  {journal} {J.
  Phys. A}\ }\textbf {\bibinfo {volume} {39}},\ \bibinfo {pages} {13507}
  (\bibinfo {year} {2006})}\BibitemShut {NoStop}%
\bibitem [{Note2()}]{Note2}%
  \BibitemOpen
  \bibinfo {note} {Because $\rho $ lies on the state-set's boundary, the
  gradient of $F$ need not vanish in order for $\protect \mathaccentV
  {hat}05E{\rho }=\rho $ to be a local maximum.}\BibitemShut {Stop}%
\bibitem [{\citenamefont {Sugiyama}\ \emph {et~al.}(2012)\citenamefont
  {Sugiyama}, \citenamefont {Turner},\ and\ \citenamefont
  {Murao}}]{SugiyamaNJP12}%
  \BibitemOpen
  \bibfield  {author} {\bibinfo {author} {\bibfnamefont {T.}~\bibnamefont
  {Sugiyama}}, \bibinfo {author} {\bibfnamefont {P.~S.}\ \bibnamefont
  {Turner}}, \ and\ \bibinfo {author} {\bibfnamefont {M.}~\bibnamefont
  {Murao}},\ }\href@noop {} {\bibfield  {journal} {\bibinfo  {journal} {N. J.
  Phys.}\ }\textbf {\bibinfo {volume} {14}},\ \bibinfo {pages} {085005}
  (\bibinfo {year} {2012})}\BibitemShut {NoStop}%
\bibitem [{Note3()}]{Note3}%
  \BibitemOpen
  \bibinfo {note} {The ``$O$'' notation is necessary here because some of the
  remaining $N-N_0$ copies may be measured in other bases that make up a
  complete measurement frame.}\BibitemShut {Stop}%
\bibitem [{\citenamefont {L\"{o}fberg}(2004)}]{yalmip}%
  \BibitemOpen
  \bibfield  {author} {\bibinfo {author} {\bibfnamefont {J.}~\bibnamefont
  {L\"{o}fberg}},\ }\href@noop {} {\bibfield  {journal} {\bibinfo  {journal}
  {Proc. CACSD (Taipei)}\ } (\bibinfo {year} {2004})}\BibitemShut {NoStop}%
\bibitem [{Note4()}]{Note4}%
  \BibitemOpen
  \bibinfo {note} {All quoted uncertainties herein are $1\sigma $, or 68\%
  confidence intervals. Therefore, we don't expect the the ``true'' value to
  lie within the error bars more than 68\% of the time. Most of the results
  given here agree with theoretical predictions to within $2\sigma $ (95\%
  confidence intervals), a common criterion for consistency between data and
  theory.}\BibitemShut {Stop}%
\bibitem [{\citenamefont {James}\ \emph {et~al.}(2001)\citenamefont {James},
  \citenamefont {Kwiat}, \citenamefont {Munro},\ and\ \citenamefont
  {White}}]{JamesPRA01}%
  \BibitemOpen
  \bibfield  {author} {\bibinfo {author} {\bibfnamefont {D.~F.~V.}\
  \bibnamefont {James}}, \bibinfo {author} {\bibfnamefont {P.~G.}\ \bibnamefont
  {Kwiat}}, \bibinfo {author} {\bibfnamefont {W.~J.}\ \bibnamefont {Munro}}, \
  and\ \bibinfo {author} {\bibfnamefont {A.~G.}\ \bibnamefont {White}},\ }\href
  {\doibase 10.1103/PhysRevA.64.052312} {\bibfield  {journal} {\bibinfo
  {journal} {Phys. Rev. A}\ }\textbf {\bibinfo {volume} {64}},\ \bibinfo
  {pages} {052312} (\bibinfo {year} {2001})}\BibitemShut {NoStop}%
\bibitem [{\citenamefont {H{\"u}bner}(1992)}]{HubnerPLA92}%
  \BibitemOpen
  \bibfield  {author} {\bibinfo {author} {\bibfnamefont {M.}~\bibnamefont
  {H{\"u}bner}},\ }\href@noop {} {\bibfield  {journal} {\bibinfo  {journal}
  {Phys. Lett. A}\ }\textbf {\bibinfo {volume} {163}},\ \bibinfo {pages} {239}
  (\bibinfo {year} {1992})}\BibitemShut {NoStop}%
\bibitem [{\citenamefont {Petz}\ and\ \citenamefont {Sudar}(1996)}]{PetzJMP96}%
  \BibitemOpen
  \bibfield  {author} {\bibinfo {author} {\bibfnamefont {D.}~\bibnamefont
  {Petz}}\ and\ \bibinfo {author} {\bibfnamefont {C.}~\bibnamefont {Sudar}},\
  }\href@noop {} {\bibfield  {journal} {\bibinfo  {journal} {J. Math. Phys}\
  }\textbf {\bibinfo {volume} {37}},\ \bibinfo {pages} {2662} (\bibinfo {year}
  {1996})}\BibitemShut {NoStop}%
\bibitem [{\citenamefont {\ifmmode~\dot{Z}\else \.{Z}\fi{}yczkowski}\ and\
  \citenamefont {Sommers}(2005)}]{ZyczkowskiPRA05}%
  \BibitemOpen
  \bibfield  {author} {\bibinfo {author} {\bibfnamefont {K.}~\bibnamefont
  {\ifmmode~\dot{Z}\else \.{Z}\fi{}yczkowski}}\ and\ \bibinfo {author}
  {\bibfnamefont {H.-J.}\ \bibnamefont {Sommers}},\ }\href@noop {} {\bibfield
  {journal} {\bibinfo  {journal} {Phys. Rev. A}\ }\textbf {\bibinfo {volume}
  {71}},\ \bibinfo {pages} {032313} (\bibinfo {year} {2005})}\BibitemShut
  {NoStop}%
\bibitem [{Note5()}]{Note5}%
  \BibitemOpen
  \bibinfo {note} {Ironically, restricting the problem to pure states falsely
  trivializes it -- the average \protect \emph {and} worst-case infidelity is
  $O(1/N)$ even for static tomography! The difficulty is not in estimating
  which pure state we have, but in distinguishing between small eigenvalues
  ($\lambda =0$ vs $\lambda =1/\protect \sqrt {N}$).}\BibitemShut {Stop}%
\bibitem [{\citenamefont {Massar}\ and\ \citenamefont
  {Popescu}(1995)}]{MassarPRL95}%
  \BibitemOpen
  \bibfield  {author} {\bibinfo {author} {\bibfnamefont {S.}~\bibnamefont
  {Massar}}\ and\ \bibinfo {author} {\bibfnamefont {S.}~\bibnamefont
  {Popescu}},\ }\href {\doibase 10.1103/PhysRevLett.74.1259} {\bibfield
  {journal} {\bibinfo  {journal} {Phys. Rev. Lett.}\ }\textbf {\bibinfo
  {volume} {74}},\ \bibinfo {pages} {1259} (\bibinfo {year}
  {1995})}\BibitemShut {NoStop}%
\end{thebibliography}%
\section{Supplementary Material}

In our paper, we attribute certain properties of our experimental data to ``systematic errors''.  The purpose of this Supplement is to provide more detail on the role that systematic errors play in our experimental results, and their interplay with static and adaptive tomography.  ``Systematic errors'' is a broad term, incorporating almost everything that can go wrong with an experiment, so we consider several forms of it.  We begin by briefly discussing \emph{frame misalignment}, where adaptive tomography yields no advantage, but almost nothing else can either.  We then consider some systematic errors in measurement that can be detected and mitigated, and demonstrate through simulations that they affect static tomography and adaptive tomography differently.   Using a one-parameter fit and a model of our experiment, we show that waveplate-alignment errors of around $1.5\times10^{-3}$ radians reproduce our experimental results remarkably well.  To wrap up, we examine the asymptotic scaling of tomographic infidelity for three different models of systematic error, and conclude that adaptive tomography mitigates these forms of systematic error much better than static tomography can.

\textbf{Frame Misalignment}:  The most systematic of errors is a fixed misalignment of reference frames.  In our linear optics experiment, where a set of waveplates and a polarizing beamsplitter are used to measure the polarization state of light, this means that all the optical elements are misaligned by the same amount.  Varying tomographic strategies can have no effect on this sort of error.  In fact, this kind of frame misalignment cannot be detected at all within the experiment.  It is equivalent to a change of gauge, has no operational consequences in this context, and is not interesting.

Instead, let us consider some errors that, while less ``systematic'' than frame misalignment, are also more detectable -- and therefore potentially sensitive to different tomographic strategies.  

\begin{figure}[ht]
\includegraphics[width=\FCW]{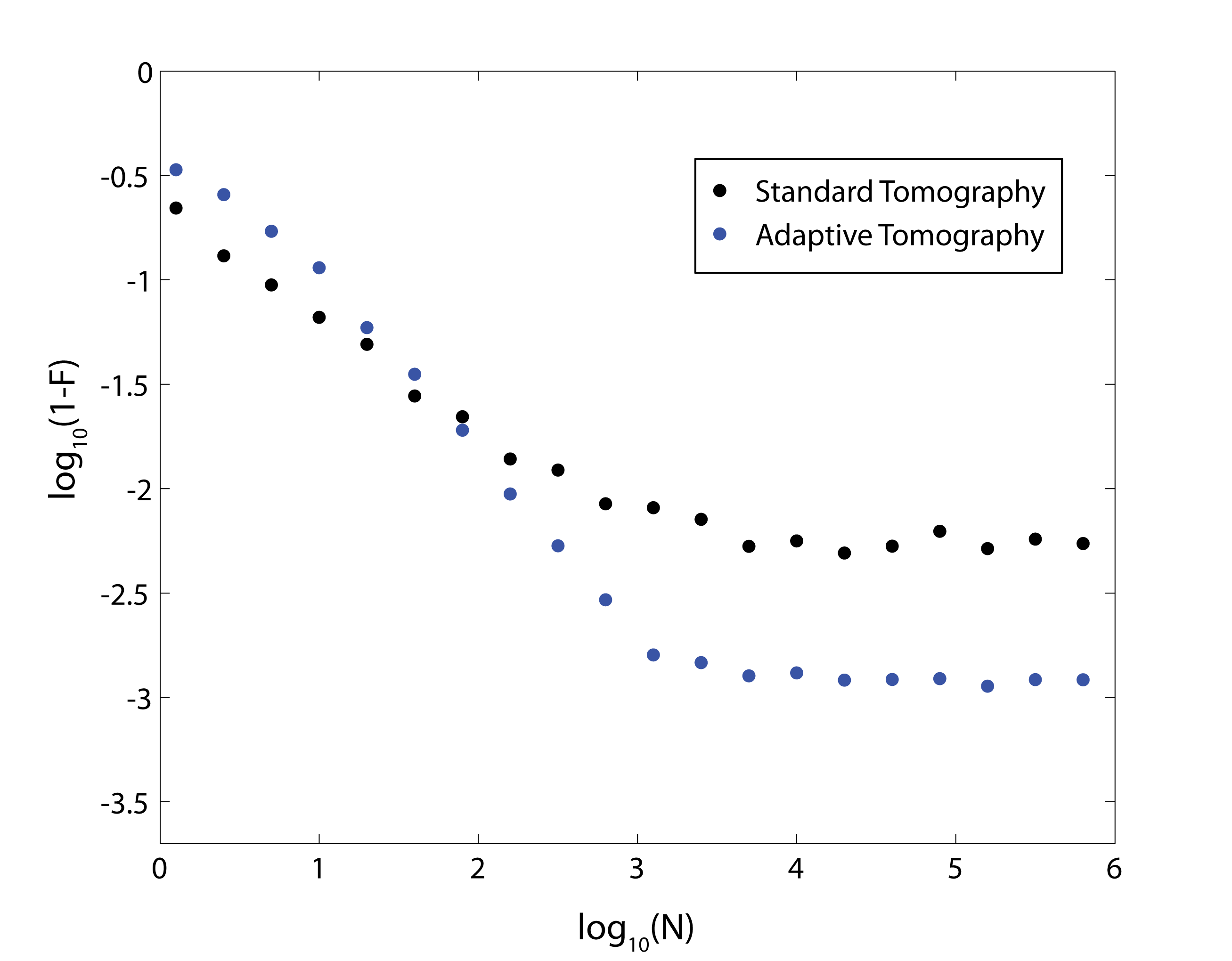}
\caption{ Average infidelity vs. sample size ($N$) for simulations with systematic errors (of Model 1 type; see text) on the order of $E=10^{-2}$.  The infidelity decreases with increasing $N$ up to a point, after which it flattens out after hitting a 'noise floor'.  The noise floor occurs at a lower average infidelity for adaptive tomography than for static tomography. \label{fig11}}
\end{figure}

\textbf{Waveplate Misalignment}:  A very important source of errors for our experiment is the alignment angle of the waveplates that measure  photon polarization states.  Our experiment's reference frame is defined by the polarizing beam-splitter used to make the final projective measurement.  To change the basis of the measurement, waveplates are mechanically rotated by a motor with good but finite accuracy.  Every time we change the measurement basis, the motor's finite accuracy causes a slight misalignment of the waveplates -- and the ensuing photodetections correspond to a measurement of a basis slightly different from the one we intended to measure.

Our terminology for discussing these errors is as follows.  We performed (and, in this Supplement, we simulate) a large number of \emph{experiments}.  A single experiment (or \emph{experimental run}) comprises the production of $N$ identically prepared photons.  Within an experiment, we implement several (3 to 6) \emph{measurement settings}.  Each measurement setting corresponds to (i) adjusting the waveplate, then (ii) measuring a large number of photons without any adjustments.  In static tomography a single experiment includes three measurement settings (projections onto X,Y, and Z axes of the Bloch sphere), in each of which $N/3$ samples are measured.  In adaptive tomography, a single experiment includes six measurement settings (the same initial set of three, and then three more in a rotated frame), each applied to $N/6$ states. 

We consider three different models of systematic error in waveplate alignment.
\begin{enumerate}
\item \textbf{Model 1}.  Each time a waveplate (whose purpose is to make a measurement) is moved to a new angle $\theta$, it ends up instead aligned at angle $\theta + \delta\theta$, where $\delta\theta$ is a Gaussian random variable with zero mean and standard deviation $E$.  Thus, in each experimental run, each measurement setting is misaligned by an independent random angle.  This angle persists over many samples in the same experimental run, but not across multiple experimental runs.  
\item \textbf{Model 2}.  Each time a waveplate is moved, it misses its target $\theta$ by a random angle $\delta\theta$ that is fixed for each \emph{experiment}, rather than for each measurement setting within the experiment.  This model is mathematically equivalent to (and can be taken to represent) a small misalignment of the polarizing beam splitter.  We take $\delta\theta$, which is fixed for each individual experiment, to be a Gaussian random variable with zero mean and standard deviation $E$.
\item \textbf{Model 3}.  The waveplates are misaligned by an angle $\delta\theta$ that is fixed for each experiment (as in Model 2), but each experiment has the same fixed misalignment.  In this model, $\delta\theta = E$ is not a random variable.
\end{enumerate}
We believe that Model 1 best represents our experiment. The waveplate motors used in our experiment have an finite precision, and every time we change their angle, they return to a factory-set ``home'' position before realigning (which eliminates or at least minimizes correlation between successive alignment errors).  Thus, the waveplate angle picks up a different random error each time it is moved to a new measurement setting.  

We simulated the effect of Models 1-3 on adaptive and static tomography.  In each simulation, 200 independent experimental runs were generated (each involving at least 3 measurement settings, with many identically prepared photons measured at each setting).  We averaged the tomographic infidelity of these 200 runs to characterize the effect of random waveplate errors in each model.  

\begin{figure}[ht]
\includegraphics[width=\FCW]{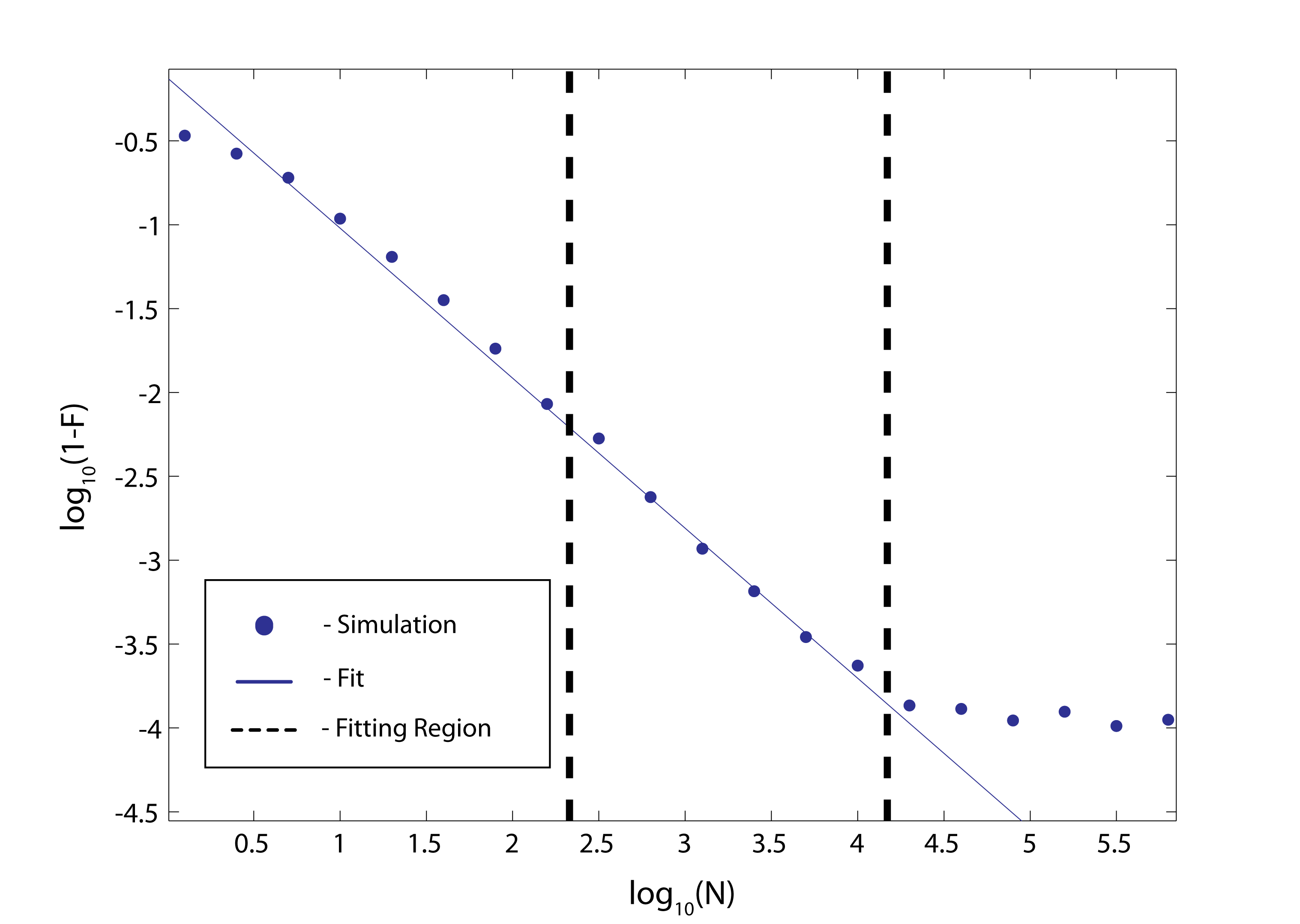}
\caption{Average infidelity vs. sample size ($N$) for simulations of adaptive tomography with systematic errors (of Model 1 type; see text) on the order of $E=10^{-3}$.  Also plotted is the region over which experimental data was taken (see main body of text) and a line of best fit for this region. \label{fig22}}
\end{figure}

\textbf{Results:}
Figure \ref{fig11} shows error (average infidelity) versus sample size ($N$) for Model 1 with $E=0.5$ degrees ($\sim 10^{-2}$ radians).  As the sample size increases, statistical errors decrease, and so average infidelity decreases.  However, the error reaches a clear noise floor as systematic errors begin to dominate.  It is higher for standard tomography than for adaptive tomography.  We conclude that adaptive tomography is less sensitive than standard tomography to systematic errors.

Since the alignment errors vary randomly from experiment to experiment, an astute experimentalist might achieve higher accuracy by repeating each measurement setting many ($M$) times, resetting the waveplate each time.  This would work, reducing the infidelity by a factor of $1/\sqrt{M}$, but it adds significantly to the experimental difficulty and complexity.  For example, if the waveplates have a precision of $\sim0.5$ degrees, then when these systematic errors dominate over statistical errors (see Figure \ref{fig11}), the infidelity of static tomography saturates at $10^{-2}$, while for adaptive tomography it saturates at $10^{-3}$.  Achieving the same $10^{-3}$ accuracy with static tomography would require repeating each measurement $M=100$ times.  Or, the experimentalist could just use adaptive tomography, and achieve it with only a single extra waveplate setting.

\begin{figure}[ht!]
\includegraphics[width=\FCW]{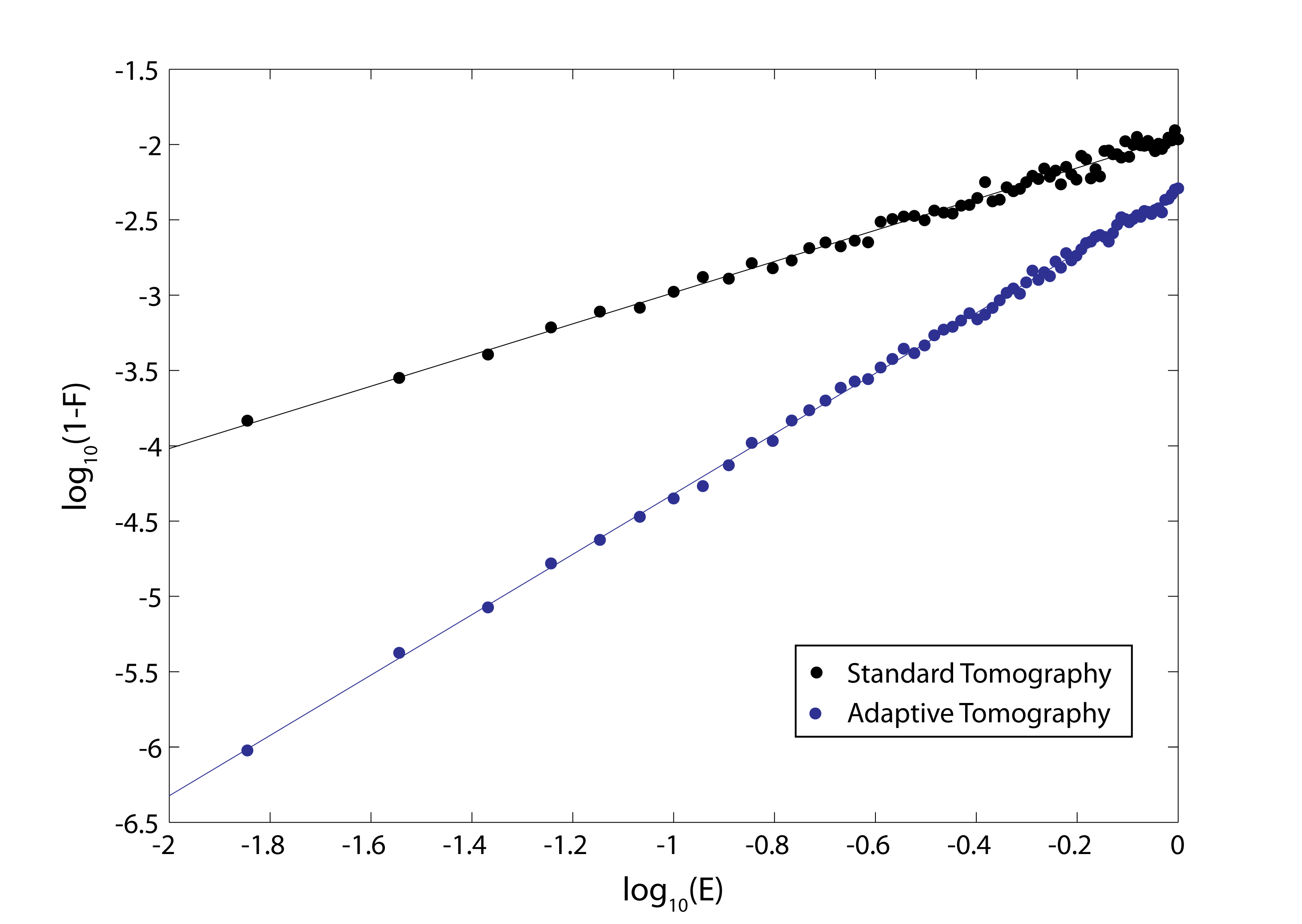}
\includegraphics[width=\FCW]{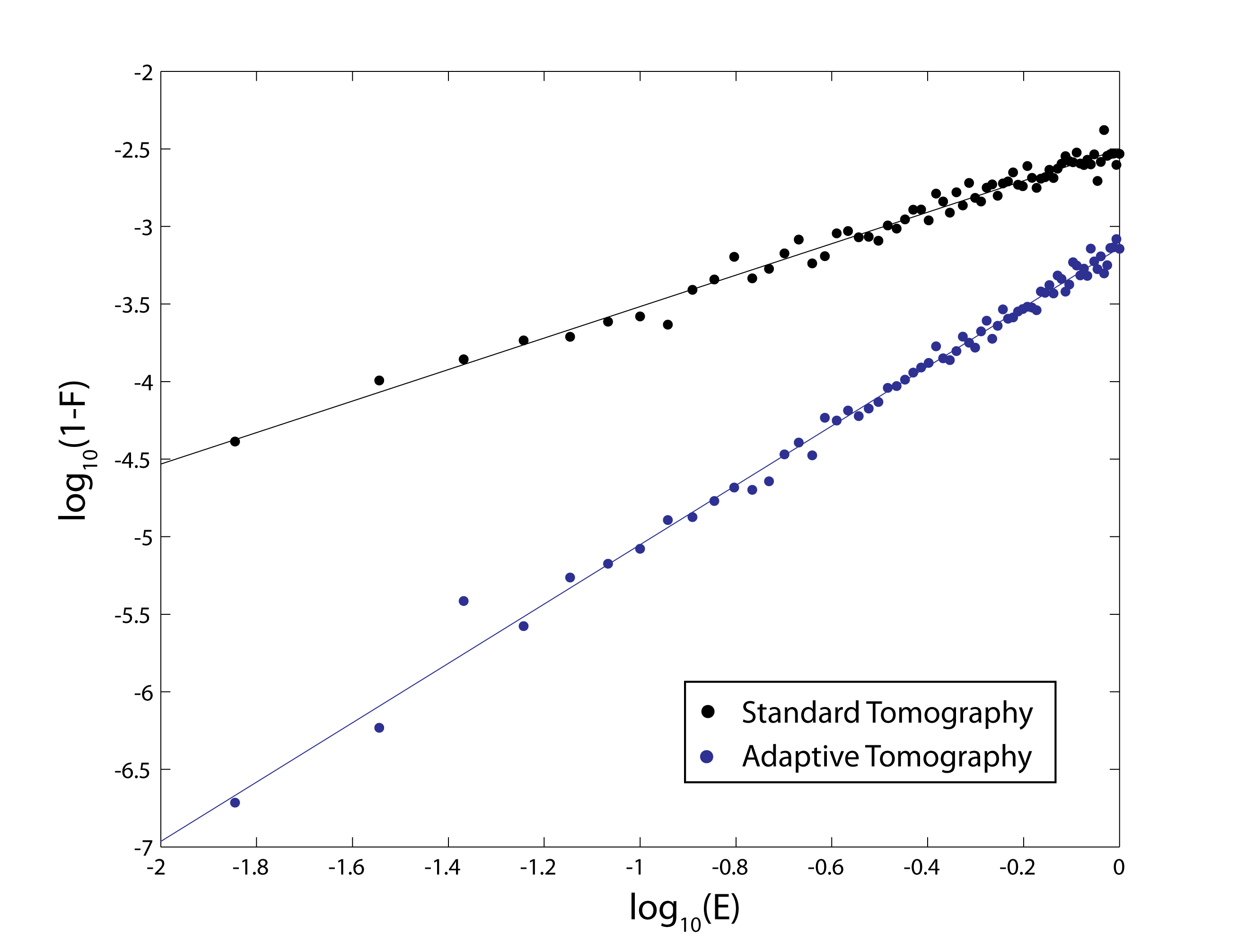}
\includegraphics[width=\FCW]{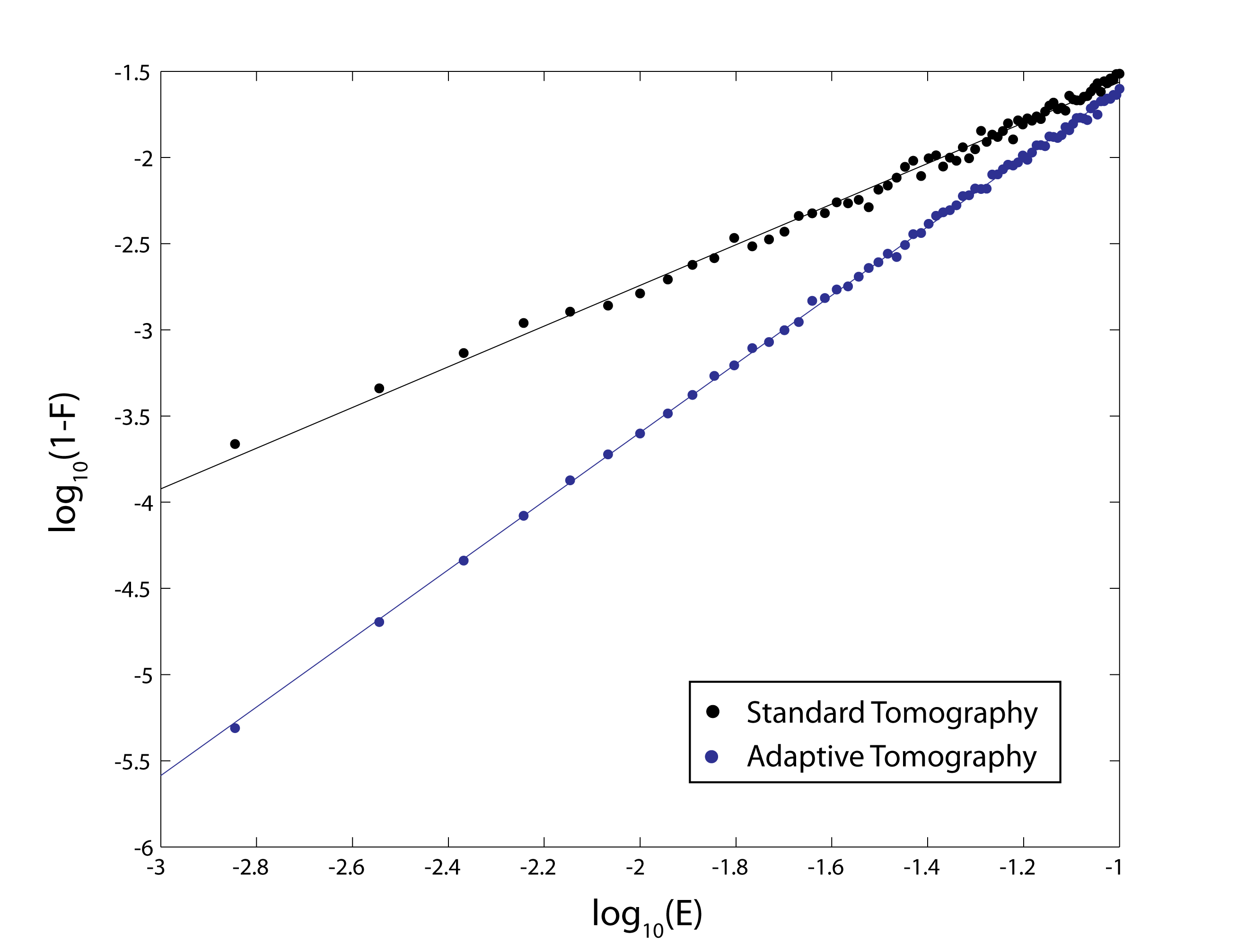}
\caption{\textbf{Location of the noise floor ($1-F$) for Models 1,2,3 (from top to bottom).}  For each of the three Models, we plot the average infidelity as $N\to\infty$ (to ensure that systematic errors dominate) vs. $E$ (the magnitude of systematic error).  Data points are results of simulation, and lines are lines of best fit.  \label{fig33}  }
\end{figure}
  
We then performed several simulations of Model 1, in which we varied $E$ (the magnitude of systematic error).  For each value of $E$, we fit a line to the $1-F$ vs. $N$ curve \emph{over the same range of $N$ that we observed in our experiment}.  This is shown in Figure \ref{fig2}.  A simulation with $E=0.15$ degrees yielded results almost indistinguishable from our experimental data.  In the experimentally observed region, the line of best fit has a slope of $-0.895 \pm 0.023 $,
$$1-F \propto N^{-0.895 \pm 0.023},$$
which matches our experimental data very well.

Finally, we investigated the value of the noise floor for Models 1-3 (Figure \ref{fig33}).  Each plot in Figure \ref{fig33} shows $1-F$ (average infidelity) vs. $E$ (magnitude of systematic errors) on a log-log plot, for both standard and adaptive tomography.  We examined sufficiently high $N$ to guarantee that systematic errors dominate.
\begin{enumerate}
\item For Model 1, the line of best fit (on a log-log plot) for standard tomography has a slope of $1.03 \pm 0.01$ and the line of best fit for adaptive tomography has a slope of $2.00 \pm 0.01$.
\item For Model 2, the slopes of the lines of best fit are $1.01 \pm 0.02$ and $1.91\pm 0.02$.  
\item For Model \#3, the slopes of the lines of best fit are $1.18\pm0.01$ and $1.99\pm0.01$.  
\end{enumerate}
We conclude that in all three models of systematica error that we considered here, it's fair to say that average infidelity scales \emph{linearly} with $E$ for standard tomography, and \emph{quadratically} with $E$ for adaptive tomography.  Adaptive tomography is substantially more robust to systematic errors than standard tomography -- not just by a constant factor, but qualitatively so.

\textbf{Conclusion:}  We have shown that for three reasonable models of systematic errors, the average infidelity of adaptive tomography scales with $E^2$ and the average infidelity of static tomography scales with $E$, where $E$ is the magnitude of these errors.  Infidelity is very sensitive to spectral errors (i.e., changes in the eigenvalues of the estimated density matrix), but not to unitary errors (changes in the eigenvectors).  The primary result of systematic errors in the measurement basis -- i.e., measuring the wrong basis by an angle $E$ -- is a unitary error in the estimate by $O(E)$.  As we have shown in the main text, adaptive tomography measures the eigenvalues of the density matrix to much higher precision than static tomography.  Furthermore, adaptive tomography (because it specifically seeks to measure the diagonal basis of $\rho$) still obtains an accurate estimate for the eigenvalues even in the presence of systematic error.  Even if we get the basis wrong by an angle of $O(E)$, this only affects the measurement probabilities (and therefore the estimated spectrum) by $O(E^2)$.

\end{document}